# A note on Coulomb collision operator in curvilinear coordinates

P R Goncharov

Saint-Petersburg Polytechnic University, 195251, Russia

E-mail: goncharov@phtf.stu.neva.ru

**Abstract.** Dynamic friction force, diffusion tensor, flux density in velocity space, and Coulomb collision term are expressed in curvilinear coordinates via Trubnikov potential functions corresponding to each species of background plasma. For comparison, explicit formulas are given for dynamic friction force, diffusion tensor and collisional flux density in velocity space in curvilinear coordinates via Rosenbluth potential functions summed over all species of background plasma.

## 1. Introduction

The collision term of the kinetic equation in case of Coulomb interaction was obtained in [1] and its equivalent formulation was given in [2,3] in Cartesian coordinates via partial potential functions written for each species of background plasma. Monograph [4] contains correct expressions of the flux density in velocity space and the collision term, using Trubnikov potentials [2,3], in spherical polar coordinates, missing general formulas in curvilinear coordinates. The purpose of this paper is to fill this noteworthy hiatus.

Regarding Rosenbluth potential functions summed over all species of background plasma, covariant formalism was given in [5] without introducing the flux in velocity space explicitly. Straightforward formulas are given here for the flux density in velocity space with explicit dynamic friction and diffusion terms, and for the collision term as minus divergence of the flux density.

In section 2 equivalent formulations [2,3] and [5] are briefly reproduced in Cartesian coordinates. General formulas in curvilinear coordinates are obtained in section 3 for the flux density in velocity space and for the collision term expressed via either partial or summed potentials.

Let $f_\alpha(\mathbf{v})$ be velocity distribution function of test particles of type $\alpha$, and $f_\beta(\mathbf{v}')$ be velocity distribution functions of background plasma species counted by index $\beta$. Functions $f_\alpha(\mathbf{v})$ and $f_\beta(\mathbf{v}')$ are normalized to unity. Let $n_\alpha$ and $n_\beta$ denote densities of particles $\alpha$ and $\beta$, respectively, and $u = |\mathbf{v} - \mathbf{v}'|$ denote the magnitude of the relative velocity of particles $\alpha$ and $\beta$. Partial potential functions [2,3] corresponding to species $\alpha$ interaction with species $\beta$ are

$$\Phi_{\alpha\beta} = -\frac{1}{4\pi}\int\frac{n_\beta f_\beta(\mathbf{v}')}{u}d^3\mathbf{v}', \tag{1}$$

$$\Psi_{\alpha\beta} = -\frac{1}{8\pi}\int u n_\beta f_\beta(\mathbf{v}')d^3\mathbf{v}', \tag{2}$$

where the integration is over the entire velocity space. Introducing a constant

$$L_{\alpha\beta} = \frac{(4\pi Z_\alpha Z_\beta e^2)^2 \Lambda}{m_\alpha^2}, \tag{3}$$

where $Z_\alpha$ and $Z_\beta$ are the electric charge numbers of particles of species $\alpha$ and $\beta$, respectively, $e$ is the elementary charge, $\Lambda$ is Coulomb logarithm, and $m_\alpha$ is the mass of a particle of species $\alpha$, we define slightly modified potential functions for convenience,

$$\tilde{\Phi}_{\alpha\beta} = \frac{m_\alpha^2}{m_\beta} L_{\alpha\beta} \Phi_{\alpha\beta}, \tag{4}$$

$$\tilde{\Psi}_{\alpha\beta} = -L_{\alpha\beta} \Psi_{\alpha\beta}, \tag{5}$$

where $m_\beta$ is the mass of a particle of species $\beta$.

Potential functions (1) and (2) are related to potential functions summed over all species of background plasma

$$H_\alpha = \sum_\beta \frac{m_\alpha + m_\beta}{m_\beta} Z_\beta^2 \int \frac{n_\beta f_\beta(\mathbf{v}')}{u}d^3\mathbf{v}', \tag{6}$$

$$G_\alpha = \sum_\beta Z_\beta^2 \int u n_\beta f_\beta(\mathbf{v}')d^3\mathbf{v}', \tag{7}$$

proposed in [5], in the following manner:

$$H_\alpha = -4\pi\sum_\beta \frac{m_\alpha + m_\beta}{m_\beta} Z_\beta^2 \Phi_{\alpha\beta}, \tag{8}$$

$$G_\alpha = -8\pi\sum_\beta Z_\beta^2 \Psi_{\alpha\beta}. \tag{9}$$

Note that in [5] $Z_\alpha = Z_\beta = 1$. Subscripts $\alpha$ and $\beta$ are not to be confused with tensor indices.

## 2. Equivalent forms of Coulomb collision term in Cartesian coordinates

The dynamic friction force and the diffusion tensor in velocity space, ascribable to collisions of particles $\alpha$ with particles $\beta$, are expressed correspondingly as minus gradient of scalar potential (4)

$$F^i = -\frac{\partial \tilde{\Phi}_{\alpha\beta}}{\partial v^i} \tag{10}$$

and Hessian tensor associated with scalar potential (5)

$$D_{ij} = \frac{\partial^2 \tilde{\Psi}_{\alpha\beta}}{\partial v^i \partial v^j}. \tag{11}$$

Summation over repeated tensor indices is assumed hereinafter. The flux of particles $\alpha$ in velocity space due to collisions with particles $\beta$ is then

$$\gamma^j_{\alpha\beta} = \frac{F^i}{m_\alpha}(n_\alpha f_\alpha) - D_{ij}\frac{\partial(n_\alpha f_\alpha)}{\partial v^j}, \tag{12}$$

and the partial collision term due to collisions between particles of type $\alpha$ with particles of type $\beta$, as shown in [3], equals minus divergence of the flux in velocity space (12)

$$C_{\alpha\beta} = -\frac{\partial}{\partial v^i}\gamma^i_{\alpha\beta}. \tag{13}$$

The full collision term equivalent to [1] is

$$C_\alpha = \sum_\beta C_{\alpha\beta}. \tag{14}$$

To reformulate (14) via summed potential functions (6) and (7), let us first introduce another constant

$$L_\alpha = \frac{4\pi(Z_\alpha e^2)^2 \Lambda}{m_\alpha^2}, \tag{15}$$

and then simply rewrite the flux (12) of test particles $\alpha$ in velocity space due to collisions with background plasma particles $\beta$ as

$$\gamma^j_{\alpha\beta} = -4\pi L_\alpha \left(\left(\frac{m_\alpha}{m_\beta}+1-1\right)Z_\beta^2 \frac{\partial \Phi_{\alpha\beta}}{\partial v^i}(n_\alpha f_\alpha) - Z_\beta^2 \frac{\partial^2 \Psi_{\alpha\beta}}{\partial v^i \partial v^j}\frac{\partial(n_\alpha f_\alpha)}{\partial v^j}\right). \tag{16}$$

Using the identity proven in [2,3] that potential (1) equals Laplacian of potential (2)

$$\Phi_{\alpha\beta} = \frac{\partial^2 \Psi_{\alpha\beta}}{\partial v_j \partial v^j}, \tag{17}$$

we rewrite (16) once more as

$$\gamma^j_{\alpha\beta} = -4\pi L_\alpha \left( \frac{m_\alpha + m_\beta}{m_\beta} Z^2_\beta \frac{\partial \Phi_{\alpha\beta}}{\partial v^i}(n_\alpha f_\alpha) - Z^2_\beta \frac{\partial}{\partial v^i}\frac{\partial^2 \Psi_{\alpha\beta}}{\partial v_j \partial v^j}(n_\alpha f_\alpha) - Z^2_\beta \frac{\partial^2 \Psi_{\alpha\beta}}{\partial v^i \partial v^j}\frac{\partial (n_\alpha f_\alpha)}{\partial v^j} \right). \quad (18)$$

Bearing in mind (8) and (9), we obtain the summed flux of particles $\alpha$ in velocity space due to collisions with all background plasma species $\beta$

$$\gamma^j_\alpha = \sum_\beta \gamma^j_{\alpha\beta} = L_\alpha \left( \frac{\partial H_\alpha}{\partial v^i}(n_\alpha f_\alpha) - \frac{1}{2}\frac{\partial}{\partial v^i}\frac{\partial^2 G_\alpha}{\partial v_j \partial v^j}(n_\alpha f_\alpha) - \frac{1}{2}\frac{\partial^2 G_\alpha}{\partial v^i \partial v^j}\frac{\partial (n_\alpha f_\alpha)}{\partial v^j} \right). \quad (19)$$

This can be rewritten as a sum of dynamic friction and diffusion terms in velocity space

$$\gamma^j_\alpha = \frac{\mathcal{F}^i}{m_\alpha}(n_\alpha f_\alpha) - \mathcal{D}_{ij}\frac{\partial (n_\alpha f_\alpha)}{\partial v^j}, \quad (20)$$

where the full dynamic friction force is

$$\mathcal{F}^i = m_\alpha L_\alpha \left( \frac{\partial H_\alpha}{\partial v^i}(n_\alpha f_\alpha) - \frac{1}{2}\frac{\partial}{\partial v^i}\frac{\partial^2 G_\alpha}{\partial v_j \partial v^j}(n_\alpha f_\alpha) \right), \quad (21)$$

and the full diffusion tensor in velocity space is

$$\mathcal{D}_{ij} = \frac{1}{2}L_\alpha \frac{\partial^2 G_\alpha}{\partial v^i \partial v^j}. \quad (22)$$

The full collision term identical to (14) and also equivalent to [1] is

$$C_\alpha = -\frac{\partial}{\partial v^i}\gamma^j_\alpha. \quad (23)$$

## 3. Generalization to curvilinear coordinates

We now generalize the expressions for the dynamic friction force and the diffusion tensor in velocity space, ascribable to collisions of particles $\alpha$ with particles $\beta$. The gradient is a covariant vector. Contravariant coordinates of the dynamic friction force analogous to (10) are

$$F^i = -g^{ik}\frac{\partial \tilde{\Phi}_{\alpha\beta}}{\partial v^k}, \quad (24)$$

where $g^{ik}$ is the contravariant metric tensor. Replacing the mixed second order partial derivative in (11) by the covariant derivative of the corresponding covariant vector, we obtain the diffusion tensor in velocity space in the form of covariant Hessian tensor

$$D_{ij} = \frac{\partial^2 \tilde{\Psi}_{\alpha\beta}}{\partial v^i \partial v^j} - \Gamma_{ij}^k \frac{\partial \tilde{\Psi}_{\alpha\beta}}{\partial v^k}, \tag{25}$$

where

$$\Gamma_{ij}^k = \frac{1}{2} g^{kl} \left( \frac{\partial g_{il}}{\partial v^j} + \frac{\partial g_{jl}}{\partial v^i} - \frac{\partial g_{ij}}{\partial v^l} \right) \tag{26}$$

is a Christoffel symbol of the second kind.

Thus, we obtain contravariant coordinates of the partial flux density in velocity space

$$\gamma_{\alpha\beta}^j = \frac{F^i}{m_\alpha} (n_\alpha f_\alpha) - g^{ij} D_{jk} g^{kl} \frac{\partial (n_\alpha f_\alpha)}{\partial v^l} \tag{27}$$

and the partial collision term due to collisions between particles of type $\alpha$ with particles of type $\beta$, replacing the partial derivative in (13) by the covariant derivative

$$C_{\alpha\beta} = -\frac{\partial \gamma_{\alpha\beta}^j}{\partial v^i} - \Gamma_{ik}^i \gamma_{\alpha\beta}^k. \tag{28}$$

The full collision term is then calculated using (14).

Generalizing the expressions for gradient vector and Laplacian in (21), Hessian tensor in (22), and divergence in (23) in much the same way, we obtain contravariant coordinates of the full collisional flux density in velocity space

$$\gamma_\alpha^j = \frac{\mathcal{F}^i}{m_\alpha} (n_\alpha f_\alpha) - g^{im} \mathcal{D}_{mk} g^{kl} \frac{\partial (n_\alpha f_\alpha)}{\partial v^l}, \tag{29}$$

where

$$\mathcal{F}^i = m_\alpha L_\alpha \left( g^{im} \frac{\partial H_\alpha}{\partial v^m} - \frac{1}{2} g^{im} \frac{\partial}{\partial v^m} \left( \frac{1}{\sqrt{g}} \frac{\partial}{\partial v^j} \left( \sqrt{g} g^{jk} \frac{\partial G_\alpha}{\partial v^k} \right) \right) \right) \tag{30}$$

is the full dynamic friction force, $g$ is the determinant of the metric,

$$g = |g_{ij}|, \tag{31}$$

and

$$\mathcal{D}_{mk} = \frac{1}{2} L_\alpha \left( \frac{\partial^2 G_\alpha}{\partial v^m \partial v^k} - \Gamma_{mk}^j \frac{\partial G_\alpha}{\partial v^j} \right) \tag{32}$$

is the full diffusion tensor in velocity space. Finally, the full collision term in this notation is

$$C_\alpha = -\frac{\partial \gamma_\alpha^j}{\partial v^i} - \Gamma_{ik}^i \gamma_\alpha^k. \tag{33}$$

## 4. Spherical polar coordinates

For the case of spherical polar coordinates $(v, \vartheta, \varphi)$, where $v \in [0, +\infty)$, $\vartheta \in [0, \pi]$, and $\varphi \in [0, 2\pi)$, we have $v_x = v \sin\vartheta \cos\varphi$, $v_y = v \sin\vartheta \sin\varphi$, and $v_z = v \cos\vartheta$. The components of the covariant metric tensor are

$$g_{ij} = \begin{pmatrix} 1 & 0 & 0 \\ 0 & v^2 & 0 \\ 0 & 0 & v^2 \sin^2\vartheta \end{pmatrix}, \tag{34}$$

its determinant

$$g = v^4 \sin^2\vartheta, \tag{35}$$

and Christoffel symbols of the second kind are

$$\Gamma^v_{ij} = \begin{pmatrix} 0 & 0 & 0 \\ 0 & -v & 0 \\ 0 & 0 & -v\sin^2\vartheta \end{pmatrix}, \quad \Gamma^\vartheta_{ij} = \begin{pmatrix} 0 & 1/v & 0 \\ 1/v & 0 & 0 \\ 0 & 0 & -\sin\vartheta\cos\vartheta \end{pmatrix}, \quad \Gamma^\varphi_{ij} = \begin{pmatrix} 0 & 0 & 1/v \\ 0 & 0 & \cot\vartheta \\ 1/v & \cot\vartheta & 0 \end{pmatrix}. \tag{36}$$

In this particular case the collisional flux density and the collision term calculated using (27) and (28) coincide with the expressions given in [4]. Substituting (34)-(36) into formulas (29)-(33) leads to the result equivalent to spherical polar case in [5].

## 5. Summary

Explicit formulas have been obtained in curvilinear coordinates for dynamic friction force, diffusion tensor, flux density in velocity space and Coulomb collision term via both partial potential functions [2,3] written for each species of background plasma and mixed potential functions [5] summed over all species of background plasma. Potentials [5] are also mixed in the sense that the expression for the dynamic friction force contains both of them. Flux density in velocity space was not introduced explicitly in [5]. Partial potentials appear to be more practical, as mentioned in [3,4], which contain results in Cartesian and spherical polar systems. This note extends form [3,4] of Coulomb collision operator to arbitrary curvilinear coordinates.

## Acknowledgements

This work was partially supported by RFBR grant No. 09-02-13608-офи_ц. The author would like to express his gratitude to Prof. Y.N. Dnestrovskii and Prof. B.V. Kuteev for a favourable discussion of the manuscript.